\documentclass{elsart}

\usepackage{amssymb}
\usepackage{graphicx}

\begin{document}

\begin{frontmatter}

\title{Signature inversion -- manifestation of drift of the rotational
axis in triaxial nuclei}


\author[1]{Zao-Chun Gao},
\author[1,2]{Y. S. Chen},
\author[3]{Yang Sun}

\address[1]{China Institute of Atomic Energy, P.O. Box
275(18), Beijing 102413, P.R. China}
\address[2]{Institute of Theoretical Physics, Academia Sinica, Beijing
100080, P.R. China}
\address[3]{Department of Physics and Joint Institute for Nuclear
Astrophysics, University of Notre Dame, Notre Dame, Indiana 46556,
USA}

\begin{abstract}
% Text of abstract
A possible scheme of realizing shell model calculations for heavy
nuclei is based on a deformed basis and the projection technique.
Here we present a new development for odd-odd nuclei, in which one
starts with triaxially-deformed multi-quasi-particle
configurations, builds the shell-model space through exact
three-dimensional angular-momentum-projection, and diagonalizes a
two-body Hamiltonian in this space. The model enables us to study
the old problem of signature inversion from a different view. With
an excellent reproduction of the experimental data in the mass-130
region, the results tend to interpret the phenomenon as a
manifestation of dynamical drift of the rotational axis with
presence of axial asymmetry in these nuclei.
\end{abstract}

\end{frontmatter}

\newcommand{\epsfigbox}[5]{%
\begin{figure} \vspace{#3}%
\includegraphics[width=8cm]{#2}%
\caption{
\label{fig:#1} #5}
\vspace{#4}
\end{figure}}

\newcommand{\epsfigpox}[5]{%
\begin{figure} \vspace{#3}%
\includegraphics[width=6.6cm]{#2}%
\caption{ \label{fig:#1} #5} \vspace{#4}
\end{figure}}

%%  \epstblbox
%%  \epstblbox{tb 1}{Table}{0 pt}{0pt}{Caption}{footnote}
%%   #1 = {label}
%%   #2 = {name of fig in picture file}
%%   #3 = {extra space above}
%%   #4 = {extra space below}
%%   #5 = {caption}
%%   #6 = {footnote}

\newcommand{\epstblbox}[6]{%
\begin{table}\vspace{#3}%
\caption{
\label{tab:#1}#5}
\includegraphics{#2}
\begin{flushleft}\vspace*{-4pt}#6
\end{flushleft}
\vspace{#4}
\end{table}}

% symbol definitions:
\newcommand{\tsub}[1]{_{\mbox{\scriptsize#1}}}
\newcommand{\tsup}[1]{^{\mbox{\scriptsize#1}}}
\newcommand{\quarterthin}{\kern 0.0417em}
\newcommand{\comm}[2]{[ \quarterthin #1 , #2 \quarterthin ]}
\newcommand{\bra}[1]{\langle#1|}
\newcommand{\ket}[1]{|#1\rangle}
\newcommand{\ev}[1]{\langle#1\rangle}
\newcommand{\mel}[3]{\bra{#1}#2\ket{#3}}
\newcommand{\thin}{\thinspace}
\newcommand{\flush}{&&}

There have been many unusual and interesting features discovered
in the nuclear high-spin rotational spectra. To describe them, it
is not feasible to apply the conventional shell models constructed
in a spherical basis. Description of heavy, deformed nuclei has
therefore relied mainly on the mean-field approximations
\cite{RS80}, or sometimes on the phenomenological particle-rotor
model \cite{BMbook}. However, there has been an increasing number
of compelling evidences indicating that correlations beyond the
mean-field level are important and a proper quantum-mechanical
treatment for nuclear states is necessary. It is thus important to
develop alternative types of nuclear structure model that can
incorporate the missing many-body correlations and make
shell-model calculations possible also for heavy nuclei.

This Letter reports on a new theoretical development for
doubly-odd systems, in line with the effort of developing shell
models using deformed bases \cite{PSM,CG01,SW03,Shei99}. The
present model employs a triaxially deformed (or $\gamma$ deformed)
basis, constructs the model space by including
multi-quasi-particle (qp) states (up to 6-qp), and performs exact
three-dimensional angular momentum projection. A two-body
Hamiltonian is then diagonalized in this space. The idea has been
applied so far only in the simplest case with a
triaxially-deformed qp vacuum state \cite{Shei99,Shei01,Sun02}.
The current work is the first attempt to realize the triaxial
projected shell model idea in a realistic situation with a
configuration mixing in a multi-qp model space. As the first
application, we address the long-standing question on signature
inversion, a phenomenon which has been widely observed in nuclear
rotational spectrum but not been convincingly explained.

This phenomenon has been suggested to relate to one of the
intrinsic symmetries in nuclei, which corresponds to ``deformation
invariance" \cite{Bohr76}. Due to this property, rotational
energies $E(I)$ ($I$: total spin of a state) of a high-$j$ band
can be split into two branches with $\Delta I = 2$, classified by
the signature quantum number, $\alpha$ \cite{BMbook}. As a rule,
the energetically favored sequence has $I=j$ (mod 2) and unfavored
one $I=j+1$ (mod 2), with $j$ being, for a doubly-odd system, the
sum of the angular momenta that the last neutron and the last
proton carry. A more general signature rule from a
quantum-mechanical derivation was given in Ref. \cite{Hara91}. The
critical observation for signature inversion \cite{Exp82} is that
at low spins, the energetically unfavored sequence is abnormally
shifted downwards, exhibiting, in an $E(I)-E(I-1)$ plot, a
reversed zigzag phase to what the signature rule predicts (see
Fig. 1 below). Only beyond a moderate spin $I_{\rm{rev}}$, which
is called reversion spin \cite{UT02}, is the normal zigzag phase
restored. The cause of signature inversion has been a major
research subject for many years.

As the first attempt of explanation, triaxiality in the nuclear
shape was suggested to be the primary reason \cite{Ben84}. With
presence of $\gamma$ deformation $(0^\circ\leq\gamma\leq
60^\circ)$, the lengths of the two principal axes, the $x$- and
$y$-axis, are different. If one assumes as usual that the moment
of inertia has the same shape dependence as that of irrotational
flow, a nucleus prefers to rotate around its intermediate-length
principal axis, the $y$-axis \cite{BMbook}. However, a nucleus
with reversed zigzag phase requires a rotation around its shortest
principal axis, the $x$-axis. To describe signature inversion, one
had to introduce \cite{HM83,Ikeda89} the concept of
$\gamma$-reversed moment-of-inertia in which one changes the
rotation axis by hand. Unsatisfied with this kind of approach,
Tajima \cite{Taji94} suggested that other ingredients in addition
to triaxiality have to be taken into account. The most popular one
discussed in the literature is the neutron-proton interaction
\cite{Semm90}.

We show that the phenomenon can be naturally described by
shell-model-type calculations without invoking unusual
assumptions. We first outline our model. The wave-function can be
written as
\begin{equation}
\left|\Psi^{\sigma}_{IM}\right> = \sum _{K \kappa}
f^{\sigma}_{IK_\kappa}\,\hat P^I_{MK}\left|\Phi_\kappa \right> ,
\label{wf}
\end{equation}
in which the projected multi-qp states span the shell model space.
In Eq. (\ref{wf}), $\left|\Phi_\kappa\right>$ represents a set of
2-, 4-, and 6-qp states associated with the triaxially deformed qp
vacuum $\left|0\right>$
\begin{eqnarray}
& &\{ \alpha^\dagger_{\nu_1} \alpha^\dagger_{\pi_1}
\left|0\right>, ~~
 \alpha^\dagger_{\nu_1} \alpha^\dagger_{\nu_2}
\alpha^\dagger_{\nu_3} \alpha^\dagger_{\pi_1} \left|0\right>, ~~
\alpha^\dagger_{\nu_1} \alpha^\dagger_{\pi_1}
\alpha^\dagger_{\pi_2} \alpha^\dagger_{\pi_3} \left|0\right>,
\nonumber\\
& & \alpha^\dagger_{\nu_1} \alpha^\dagger_{\nu_2}
\alpha^\dagger_{\nu_3} \alpha^\dagger_{\pi_1}
\alpha^\dagger_{\pi_2} \alpha^\dagger_{\pi_3} \left|0\right> \}.
\label{basis}
\end{eqnarray}
The dimension in Eq. (\ref{wf}) is $(2I+1)\times n(\kappa)$, where
$n(\kappa)$ is the number of configurations and is typically in
the order of $10^2$. $\hat P^I_{MK}$ is the three-dimensional
angular-momentum-projection operator \cite{RS80}
\begin{equation}
\hat P^I_{MK} = {2I+1 \over 8\pi^2} \int d\Omega\,
D^{I}_{MK}(\Omega)\, \hat R(\Omega),
\end{equation}
and $\sigma$ in Eq. (\ref{wf}) specifies the states with the same
angular momentum $I$.

The triaxially deformed qp states are generated by the Nilsson
Hamiltonian
\begin{equation}
\hat H_N = \hat H_0 - {2 \over
3}\hbar\omega\epsilon_2\left(\cos\gamma\hat Q_0 - \sin\gamma{{\hat
Q_{+2}+\hat Q_{-2}}\over\sqrt{2}}\right),
\label{nilsson}
\end{equation}
where the parameters $\epsilon_2$ and $\gamma$ describe quadrupole
deformation and triaxial deformation, respectively. Three major
shells (N = 3, 4, 5) are considered each for neutrons and protons.
Paring correlations are included by a subsequent BCS calculation
for the Nilsson states.

The Hamiltonian consists of a set of separable forces
\begin{equation}
\hat H = \hat H_0 - {1 \over 2} \chi \sum_\mu \hat Q^\dagger_\mu
\hat Q^{}_\mu - G_M \hat P^\dagger \hat P - G_Q \sum_\mu \hat
P^\dagger_\mu\hat P^{}_\mu .
\label{hamham}
\end{equation}
In Eq. (\ref{hamham}), $\hat H_0$ is the spherical single-particle
Hamiltonian, which contains a proper spin-orbit force
\cite{bengtsson85}. The second term is quadrupole-quadrupole
($QQ$) interaction that includes the nn, pp, and np components.
The $QQ$ interaction strength $\chi$ is determined in such a way
that it has a self-consistent relation with the quadrupole
deformation \cite{PSM}. The third term in Eq. (\ref{hamham}) is
monopole pairing, whose strength $G_M$ (in MeV) is of the standard
form $G/A$, with $G=19.6$ for neutrons and 17.2 for protons, which
approximately reproduces the observed odd-even mass differences in
this mass region. The last term is quadrupole pairing, with the
strength $G_Q$ being proportional to $G_M$, the proportionality
constant being fixed as usual to be 0.16 for all nuclei considered
in this paper. In our model, wherever the quadrupole operator
appears, we use the dimensionless quadrupole operator (defined in
Section 2.4 of Ref. \cite{PSM}). We emphasize that no new terms in
the Hamiltonian are added and no interaction strengths are
individually adjusted in the present work to reproduce data.

To observe a sizable effect of signature, one important condition
is that nucleons near the Fermi levels occupy the lower part of
high-$j$ shells having smaller $K$-components. A recent summary
for the observed $\pi h_{11/2}\nu h_{11/2}$ bands in the mass-130
region has been given by Hartley {\it et al.} \cite{UT02}. Note
that experimental deduction of spins for bands in doubly-odd
nuclei is often difficult. We apply the spin values suggested by
Liu {\it et al.} \cite{Liu98} to each of the bands if spin is not
firmly determined. The normal signature rule for these bands is
that the energetically favored states have odd-integer spins
denoted by $\alpha=1$ and the unfavored ones have even-integer
spins denoted by $\alpha=0$. However, a systematic violation of
this rule has been observed in the lower spin states. In Ref.
\cite{Hara91}, a mechanism for explaining the observed signature
inversion data in some rare earth nuclei was proposed, which
employed the projected shell model \cite{PSM} based on an axially
deformed basis. This mechanism involved a crossing of two
rotational bands that have mutually opposite signature dependence.
However, early studies \cite{Unpub} showed that it is not possible
for this mechanism to explain the data in the present study
because the condition of having bands near the Fermi levels with
mutually opposite signature dependence does not appear here.

\begin{table}
\caption{Quadrupole deformation $\epsilon_2$ and triaxial
deformation $\gamma$ employed in the calculation.} \label{tab:1}
\begin{tabular}{c|ccccccc}
\hline\noalign{\smallskip}
Nucleus    & 118   & 120   & 122   & 124   & 126   & 128   & 130 \\
\noalign{\smallskip}\hline\noalign{\smallskip}
$\epsilon_2$ & 0.30  & 0.29  & 0.26  & 0.26  & 0.21  & 0.20  & 0.19 \\
$\gamma$   & 30$^\circ$ & 30$^\circ$ & 31$^\circ$ & 31$^\circ$ & 35$^\circ$ & 37$^\circ$ & 39$^\circ$ \\
\noalign{\smallskip}\hline
\end{tabular}
\end{table}

\epsfigbox{fg 1}{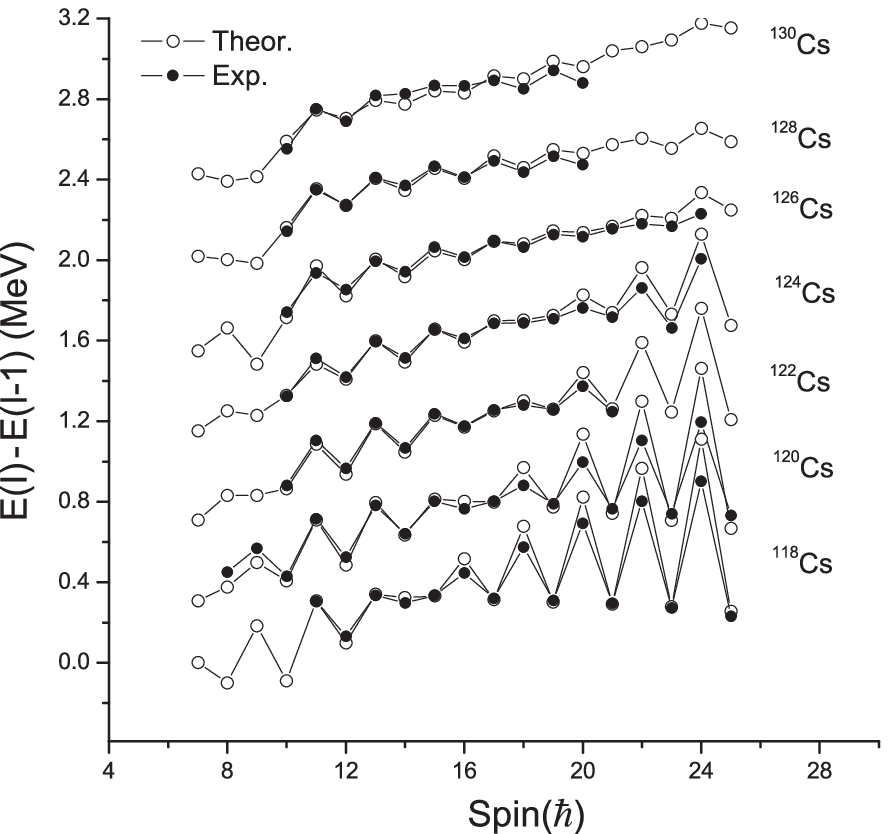}{0pt}{0pt} {Comparison of calculated
energies with data for the $\pi h_{11/2}\nu h_{11/2}$ bands in
$^{118-130}$Cs. Note the increasing trend in the reversion spin:
14.5 for $^{118}$Cs, 16.5 for $^{120}$Cs, 17.5 for $^{122}$Cs,
18.5 for $^{124}$Cs, 20.5 for $^{126}$Cs, 21.5 for $^{128}$Cs
(prediction), and 22.5 for $^{130}$Cs (prediction). Note also that
the bands are shifted vertically by $(A-118)\times 0.2$ MeV where
$A$ is mass number.}

The present calculations are performed for doubly-odd nuclei
$^{118-130}$Cs. In Table I, we list the deformation parameters
used for basis construction. The quadrupole deformation parameters
$\epsilon_2$ are consistent with those obtained from the TRS
calculations. The $\gamma$ parameters are adjusted to describe not
only the bands discussed in this paper, but also other observables
(see discussions below). Our results are compared with available
data in Fig. 1 in the form of energy difference between states of
the two signature sequences. As one can see, an excellent
agreement has been achieved. With an increasing neutron number,
the trend of decreasing signature splitting, i.e. decreasing
zigzag amplitude, has been reproduced. What is also correctly
described is the increasing trend of $I_{\rm rev}$ as neutron
number increases.

Now we investigate how signature inversion occurs in our theory.
We first study the effect of triaxiality. It has been known that
$\gamma$ is a relevant degree of freedom in this mass region, and
the nuclei are either $\gamma$-deformed or $\gamma$-soft. For
example, a stable $\gamma$ deformation is required in the
discussion of chiral doublet bands \cite{Ko03}. In Fig. 2, we take
$^{124}$Cs as an example to present the calculated bands as a
function of $\gamma$. Namely, we allow $\gamma$ in Eq.
(\ref{nilsson}) to vary, while all other parameters are fixed as
those in the $^{124}$Cs results in Fig. 1. It is observed that the
splitting between the two signature sequences is strongly
dependent on $\gamma$. With $\gamma=0$, no clear signature
splitting can be seen for the lower spin states. This explains why
one could not describe the data when an axially deformed basis is
used \cite{Unpub}. An increasing trend of splitting is obtained as
$\gamma$ increases, with the maximum splitting appearing with
$\gamma=30^\circ$. This is the $\gamma$ value corresponding to the
maximal triaxiality, after which the splitting shows a decreasing
trend. Note that signature splitting and inversion of the zigzag
phase are visible only when $\gamma$ is sufficiently large ($\ge
10^\circ$). Note also that the reversion spin $I_{\rm rev}$ does
not change with $\gamma$.

\epsfigbox{fg 2}{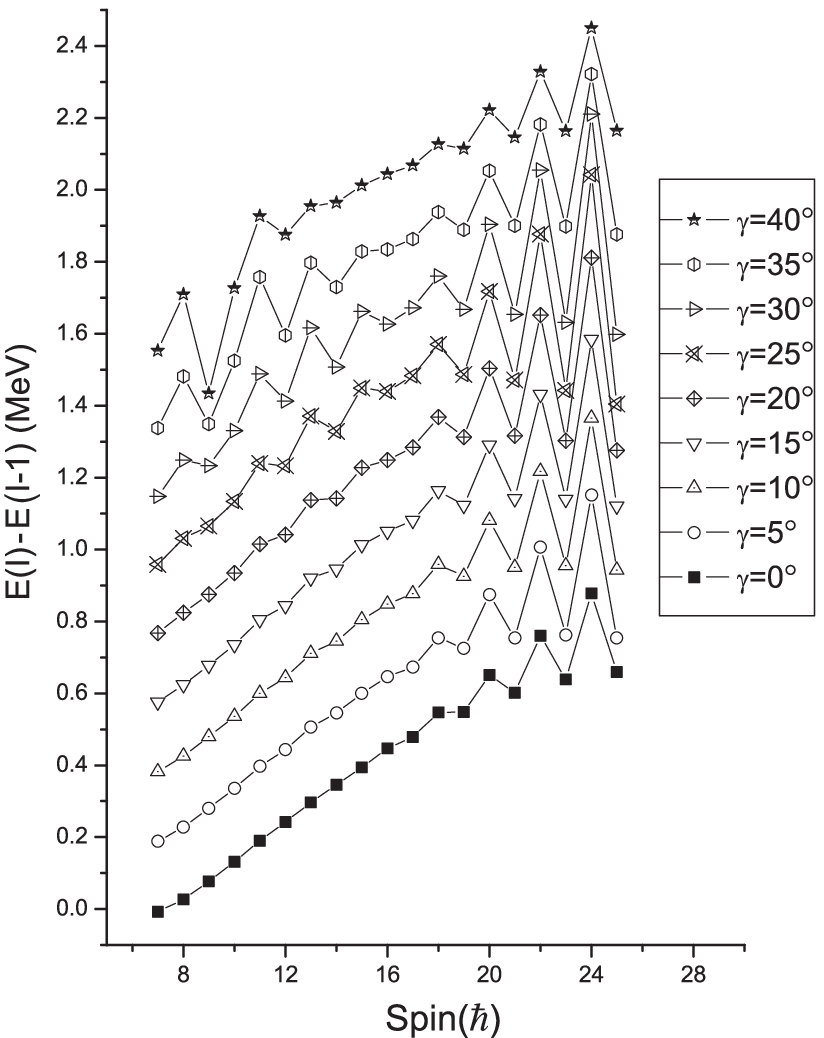}{0pt}{0pt} {Calculated 2-qp $\pi
h_{11/2}\nu h_{11/2}$ bands in $^{124}$Cs with different
triaxiality in the basis. Note that the bands are shifted
vertically by $\gamma / 25^\circ$ MeV.}

In a shell-model calculation, wavefunctions contain all
information about the evolution of a system. We demonstrate that
signature inversion and the zigzag phase restoration are the
consequence of a dynamical process in which the rotational axis
drifts from the $x$- to the $y$-direction of the principal axis as
a triaxial odd-odd nucleus is rotating. To see this, we have
calculated expectation values of the three components of
angular-momentum operator, $\hat I_i^2$ ($i = x, y$ and $z$). We
plot $I_i^2$ in Fig. 3 as functions of spin. It is seen that in
the entire spin range, $I_z^2$ takes very small, near-constant
values, indicating that the system does not rotate around the
$z$-axis. Near-zero $I_y^2$ is also seen until $I=11$; however, it
begins to climb up drastically after that spin. In contrast,
$I_x^2$ increases gradually at low spins until $I=16$, oscillates
afterwards between odd and even spins, and then quickly decreases
beyond $I=24$. We thus end up with the following picture: At low
spins, the system rotates around the $x$-axis. Starting from
$I=12$, the rotational axis begins to drift toward the
$y$-direction with a angle in the $x$-$y$ plane determined by
$I_x^2$ and $I_y^2$. For example, at $I=18$ where $I_x^2\approx
I_y^2$, the angle takes 45$^\circ$. With increasing spin, $I_y^2$
quickly dominates and the rotation eventually aligns completely
with the $y$-axis. We conclude that roughly in the spin interval
$I=12 - 24$, the rotational axis completes a drift from $x$-axis
to $y$-axis, and in this process, the reversed zigzag phase is
restored. $I_{\rm rev}$ is just the spin where the $I_x^2$ and
$I_y^2$ curves (taken average values of the odd and even spins)
cross.

\epsfigpox{fg 3}{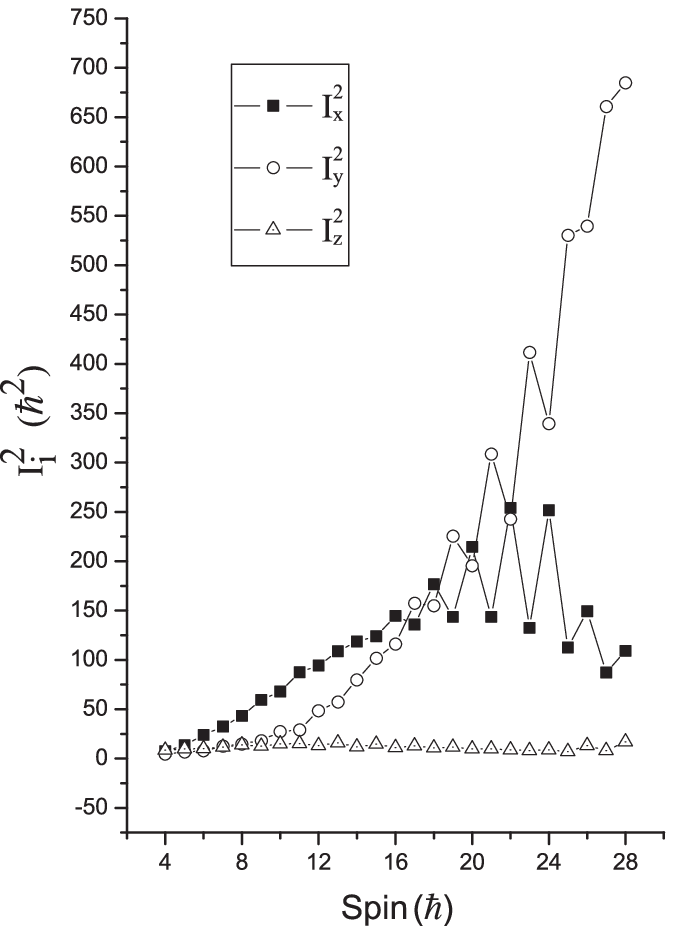}{0pt}{0pt} {Calculated expectation
values of $I_x^2$, $I_y^2$ and $I_z^2$ with the wavefunctions from
diagonalization that has reproduced the $^{124}$Cs data (see Fig.
1).}

The effect of neutron-proton interaction on the signature
inversion has been extensively discussed \cite{Semm90}. Recently,
Xu {\it et al.} have proposed \cite{Xu00} that quadrupole-pairing
force may also have an important contribution. We study the effect
of neutron-proton interaction in the $QQ$ channel by changing the
n-p interaction strength $\chi_{\rm np}$ in Eq. (\ref{hamham}). In
Fig. 4(a), we present calculations with different $\chi_{\rm np}$
by multiplying a factor $F_{\rm np}$ to its original value.
Comparing the four sets of calculations, one sees that in the
lower spin region they are nearly identical. The four curves in
the higher spin region are also very similar with differences only
in the splitting amplitude. The main effect caused by this
interaction is to shift the reversion spin. For example, with
$F_{\rm np}=0$, $I_{\rm rev}=15.5$. $I_{\rm rev}$ shifts higher to
16.5 with $F_{\rm np}=1$, and to 17.5 when $F_{\rm np}=2$. Similar
conclusion also holds for quadrupole-pairing force, as one can see
from Fig. 4(b). In the four sets of calculations with different
quadrupole-pairing strengths, the difference is only a shift of
$I_{\rm rev}$ toward higher spins with an increasing strength. We
have thus found that within our model, signature inversion occurs
with $F_{\rm np}=0$ and $G_Q=0$.

\epsfigbox{fg 4}{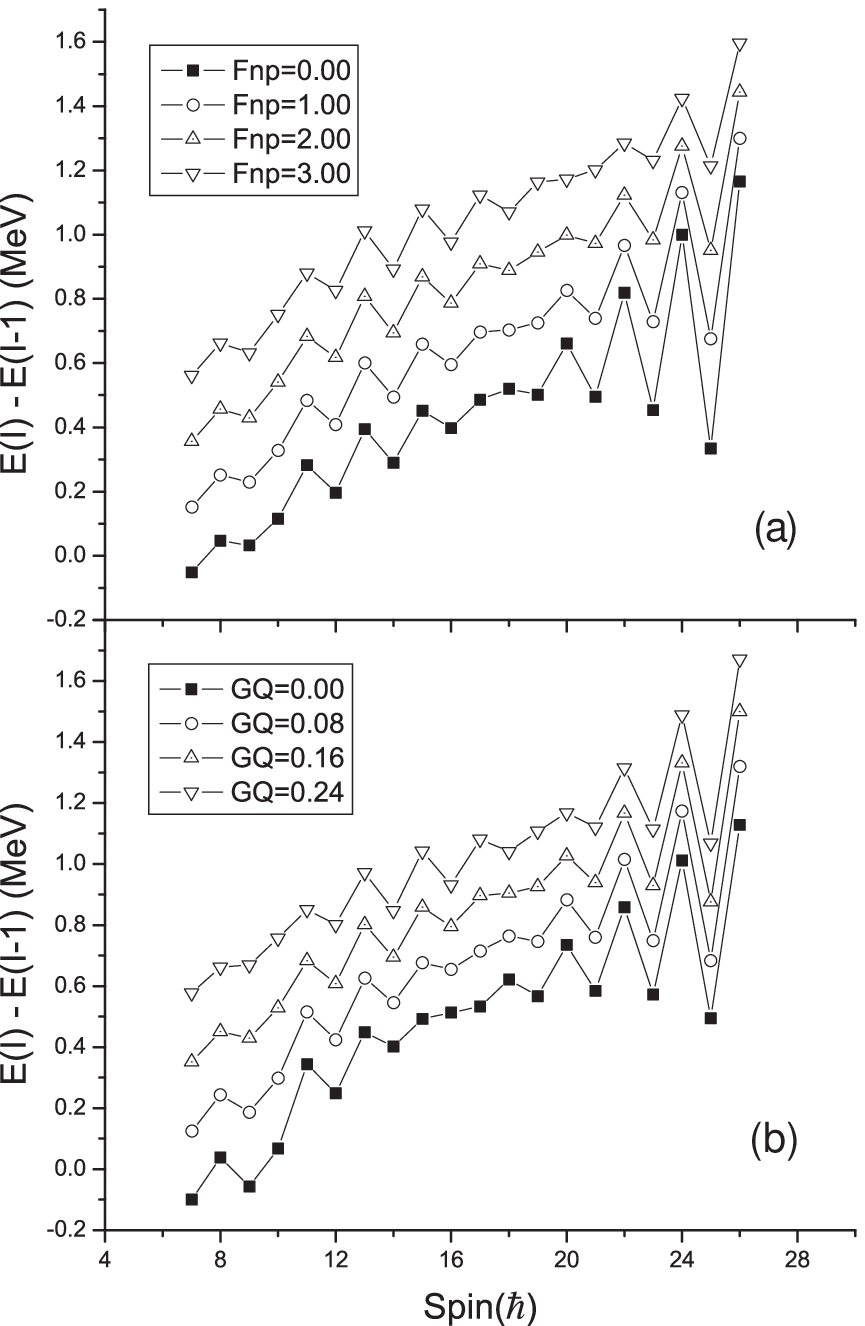}{0pt}{0pt} {Calculated $\pi h_{11/2}\nu
h_{11/2}$ bands in $^{124}$Cs with (a) different $\chi_{\rm np}$,
and (b) different $G_Q$ in the Hamiltonian (\ref{hamham}). Note
that the bands are successively shifted vertically by 0.2 MeV.}

We notice that the values of $\gamma$-parameter that account for
the experimental data of signature inversion are not always
supported by potential energy calculations from mean-field models
which in many cases predict small or zero triaxiality \cite{Xu00}.
It turns out that the correct accounting of the correlations
provided by the angular momentum projection tends to lower the
potential energy in the triaxial deformation region \cite{HHR84}.
Calculations have shown that such a tendency may be common to all
kinds of nuclei \cite{Tanabe00}. The calculated B(M1)/B(E2) in
$^{124}$Cs, as shown in Fig. 5, suggest that the $\gamma$ values
in Table I can reproduce not only the energy levels, but also the
transitions. We have verified that the same set of parameters can
well describe other observables such as the side bands in
$^{124-130}$Cs and the Yrast and $\gamma$-vibrational states in
the neighboring even-even nuclei. These results will be published
elsewhere.

\epsfigbox{fg 5}{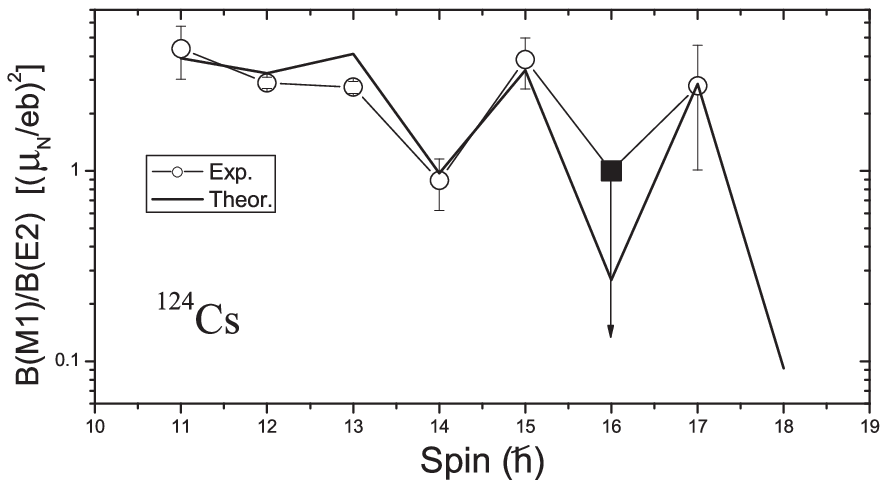}{0pt}{0pt} {Comparison of calculated
B(M1)/B(E2) with data in $^{124}$Cs.}

To summarize, the phenomenon of signature inversion has been known
for more than two decades. Here, experimental data are reproduced
systematically with high accuracy, without the need of invoking
any unusual assumptions associated with shape change or with
interactions. The key to the success may be the shell-model nature
of the method. Our states $\left| \Psi_{IM}\right>$ have an
advantage over the unprojected state $\left|0\right>$ in that the
projection treats the states fully quantum-mechanically by
collecting all the energy-degenerate mean-field states associated
with the rotational symmetry. This suggests that the effects
brought by a quantum-mechanical treatment for a nuclear system
should not be overlooked. We have shown that with the triaxial
projected shell model in a realistic configuration space,
signature inversion data in the mass-130 region can be reproduced
nicely with the separable forces in the standard form. The degree
of triaxiality in the deformed basis determines the magnitude of
signature splitting and the occurrence of signature inversion. The
residual interactions, such as the $Q_nQ_p$ and the
quadrupole-pairing force, merely modify the position where the
reversed zigzag phase is restored. By analyzing the rotational
evolution of the components of angular momentum, we have
interpreted the phenomenon of signature inversion as a
manifestation of the dynamical process in triaxial nuclei, in
which the rotational axis drifts from the shortest principal axis
to the intermediate one as nuclei are rotating. Discussion of
drift of angular momentum axis in one-quasiparticle states was
given by Ikeda and \AA berg in terms of the particle-rotor model
\cite{IS88}.

Y.-S.C. thanks M. Smith and L.L. Riedinger of ORNL, and M.
Wiescher of JINA for the warm hospitality. Y.S. thanks G.L. Long
of Tsinghua University and the Institute of Nuclear Theory at the
University of Washington for their hospitality during the
completion of this work.  Work is supported by NNSF of China under
contract No. 10305019, 10475115, 10435010, by MSBRDP of China
(G20000774), and by NSF of USA under contract PHY-0140324.

\end{document}